# AN IMPROVED ESTIMATOR IN SYSTEMATIC SAMPLING

**Rajesh Singh, Sachin Malik and Viplav Kr. Singh**
Department of Statistics, Banaras Hindu University
Varanasi-221005, India

## Abstract

In this paper we have adapted Bahl and Tuteja (1991) estimator in systematic sampling using auxiliary information. Using Bedi (1996) transformation an improved estimator is also proposed under systematic sampling. The expressions of bias and mean square error up to the first order of approximation are derived. It has been shown that the proposed estimator performs better than many existing estimators. A numerical example is included to support the theoretical results.

## 1. Introduction

There are some natural populations like forest etc., where it is not possible to apply easily the simple random sampling or other sampling schemes for estimating the population characteristics. In such situations, one can easily implement the method of systematic sampling for selecting a sample from the population. Systematic sampling has the advantage of selecting the whole sample with just random start. Estimation in systematic sampling has been discussed in detail by Lahiri (1954), Gautschi (1957), Hajeck (1959) and Cochran (1957). Use of auxiliary information in construction of estimators is considered by Kushwaha and Singh (1989), Banarasi *et al.* (1993) and Singh and Singh (1998). We introduced the following terminology to discuss the estimators.

Let y be the study variable and x be the auxiliary variable defined on a finite population $U = (U_1, U_2 .........., U_N)$. Here, we assume N = nk, where n and k are positive integers. Let $(y_{ij}, x_{ij})$; i = 1, 2,...,k; j=1, 2, ...,n denote the value of $j^{th}$ unit in the $i^{th}$ sample. The systematic sample means

$$\bar{y}^* = \frac{1}{n}\sum_{j=1}^{n} y_{ij}, \qquad \bar{x}^* = \frac{1}{n}\sum_{j=1}^{n} x_{ij}$$

are unbiased estimators of the population means $(\bar{Y}, \bar{X})$ of (y, x), respectively.

If $e_0 = \frac{(\bar{y}^* - \bar{Y})}{\bar{Y}}$ and $e_1 = \frac{(\bar{x}^* - \bar{X})}{\bar{X}}$

Then, we have

$$E(e_0) = E(e_1) = 0,$$

and

$$E(e_0^2) = \theta(1 + (n-1)\rho_y)C_y^2, \qquad E(e_1^2) = \theta(1 + (n-1)\rho_x)C_x^2$$

$$E(e_0 e_1) = \theta(1 + (n-1)\rho_y)^{\frac{1}{2}}(1 + (n-1)\rho_x)^{\frac{1}{2}}\rho C_y C_x,$$

where,

$$\theta = \frac{N-1}{Nn},$$

$$\rho_x = \frac{E(x_{ij} - \bar{X})(x_{ij} - \bar{X})}{E(x_{ij} - \bar{X})^2}, \qquad \rho_y = \frac{E(y_{ij} - \bar{Y})(y_{ij} - \bar{Y})}{E(y_{ij} - \bar{Y})^2}$$

and

$$\rho = \frac{E(x_{ij} - \bar{X})(y_{ij} - \bar{Y})}{\left(E(x_{ij} - \bar{X})^2 (y_{ij} - \bar{Y})^2\right)^{\frac{1}{2}}}.$$

$(C_y, C_x)$ are the coefficients of variation of the variates (y, x) respectively. It is assumed that the population mean $\bar{X}$ of the auxiliary variable is known. The classical ratio and product estimators for $\bar{Y}$ based on the systematic sample $(y_{ij}, x_{ij}; i = 1,2,...,k; j = 1,2,...n)$ of size n, are, respectively, defined by

$$t_1 = \left(\frac{\bar{y}^*}{\bar{x}^*}\right)\bar{X} \tag{1.1}$$

$$t_2 = \bar{y}^* \left(\frac{\bar{x}^*}{\bar{X}}\right) \tag{1.2}$$

which are respectively due to Swain (1964) and Shukla (1971).

The variances of $t_1$ and $t_2$ to the first order of approximation are, respectively, given by

$$\mathrm{Var}(t_1) = \theta \bar{Y}^2 (1 + (n-1)\rho_x)\left(\rho^{*2} C_y^2 + (1 - 2k\rho^*)C_x^2\right) \tag{1.3}$$

$$\operatorname{Var}(t_2) = \theta \bar{Y}^2 \left(1 + (n-1)\rho_x\right)\left(\rho^{*2} C_y^2 + \left(1 + 2k\rho^*\right) C_x^2\right) \tag{1.4}$$

And the variances of usual unbiased estimator $\bar{y}^*$ is given by

$$\operatorname{Var}\left(\bar{y}^*\right) = \theta \bar{Y}^2 \left(1 + (n-1)\rho_y\right) C_y^2 \tag{1.5}$$

where $$\rho^* = \frac{1 + (n-1)\rho_y}{1 + (n-1)\rho_x}, \quad k = \rho \frac{c_y}{c_x}.$$

## 2. Adapted Estimator

Bahl and Tuteja (1991) introduced an exponential ratio type estimator for population mean in simple random sampling given by

$$t_3 = \bar{y} \exp\left(\frac{\bar{X} - \bar{x}}{\bar{X} + \bar{x}}\right) \tag{2.1}$$

Motivated by Bahl and Tuteja (1991), we adept this estimator to the systematic sampling as

$$t_3 = \bar{y}^* \exp\left(\frac{\bar{X} - \bar{x}^*}{\bar{X} + \bar{x}^*}\right) \tag{2.2}$$

Expressing equation (2.2 ) in terms of e's, we have

$$t_3 = \bar{Y}(1 + e_0) \exp\left[-\frac{e_1}{2}\left(1 + \frac{e_1}{2}\right)^{-1}\right] \tag{2.3}$$

$$= \bar{Y}\left[1 + e_0 - \frac{e_1}{2} + \frac{3}{8}e_1^2 - \frac{e_0 e_1}{2}\right] \tag{2.4}$$

Subtracting $\bar{Y}$ from both the sides of equation (2.4) and then taking expectation of both sides, we get the bias of the estimator $t_3$, up to the first order of approximation, as

$$t_3 - \bar{Y} = \bar{Y}\left[e_0 - \frac{e_1}{2} + \frac{3e_1^2}{8} - \frac{e_0 e_1}{2}\right] \tag{2.5}$$

Taking expectation on both sides in (2.5) we get the bias of the estimator $t_3$ as

$$B(t_3) = \bar{Y}\theta\left[\frac{3C_1^2}{8} - \frac{C_0 C_1}{2}\right] \tag{2.6}$$

$$C_0^2 = \left(1+(n-1)\rho_y\right)C_y^2$$

where, $$C_1^2 = \left(1+(n-1)\rho_x\right)C_x^2$$

$$C_0C_1 = (1+(n-1)\rho_y)^{1/2}(1+(n-1)\rho_x)^{1/2}\rho C_y C_x \quad (2.7)$$

Squaring both sides of (2.5) and then taking expectation, we get the MSE of the estimator to the first order approximation, as

$$MSE(t_3) = \bar{Y}\theta\left[c_0^{\ 2} + \frac{c_1^{\ 2}}{4} - C_0C_1\right] \quad (2.8)$$

Following Bahl and Tuteja (1991), using a transformation proposed by Bedi (1996), Shabbir and Gupta (2011) introduced an exponential ratio type estimator for population mean in simple random sampling given by,

$$t_4 = \left[k_1\bar{y} + k_2(\bar{X}-\bar{x})\right]\exp\left(\frac{(\bar{A}-\bar{a})}{(\bar{A}+\bar{a})}\right) \quad (2.9)$$

where, $k_i \ (i=1,2)$ are constants.

Let $$a_i = x_i + N\bar{X},$$

So we have $$\bar{a} = \bar{x} + N\bar{X} \quad \text{and} \quad \bar{A} = \bar{X} + N\bar{X}$$

Motivated by Shabbir and Gupta (2011), we adapt this family to the systematic sampling as

$$t_4 = \left[k_1\bar{y}^* + k_2(\bar{X}-\bar{x}^*)\right]\exp\left(\frac{(\bar{A}-\bar{a}^*)}{(\bar{A}+\bar{a}^*)}\right) \quad (2.10)$$

where $$\bar{a}^* = \bar{x}^* + N\bar{X}\ .$$

Expressing $t_4$ in terms of $e_i$'s, we can write (2.10) as

$$t_4 = \left[k_1\bar{y}(1+e_0) - k_2\bar{X}e_1)\right]\exp\left(\frac{-e_1}{2(1+N)}\left\{1+\frac{e_1}{2(1+N)}\right\}^{-1}\right) \quad (2.11)$$

Expanding the right hand in (2.11), to the 1st order of approximation, we obtain

$$t_4 - \bar{Y} \cong \bar{Y}\left[(k_1-1) + k_1\left\{e_0 - \frac{e_1}{2(1+N)} - \frac{e_0e_1}{2(1+N)} + \frac{3e_1^{\ 2}}{8(1+N)^2}\right\}\right] - k_2\bar{X}\left\{e_1 - \frac{e_1^{\ 2}}{2(1+N)^2}\right\}$$

(2.12)

Taking expectation of both sides in (2.12), we get the bias of the estimator $t_4$ as

$$B(t_4)=\bar{Y}\left[(k_1-1)+k_1\theta\left\{\frac{3C_X{}^2}{8(1+N)^2}-\frac{C_0C_1}{2(1+N)}\right\}\right]-k_2\bar{X}\theta\frac{C_1{}^2}{2(1+N)^2} \qquad (2.13)$$

Squaring (2.12) and then taking expectation, we get the MSE of the estimator $t_4$

$$\mathrm{MSE}\,(t_4)=\bar{Y}^2\left[(k_1-1)^2+k_1{}^2\theta\left\{c_0{}^2+\frac{c_1{}^2}{(1+N)^2}-\frac{2c_1c_0}{(1+N)}\right\}-2k_1\theta\left\{\frac{3}{8}\frac{c_1{}^2}{(1+N)^2}-\frac{c_0c_1}{2(1+N)^2}\right\}\right]$$

$$+k_2{}^2\bar{X}^2C_1{}^2\theta-\frac{k_2\,\bar{X}\,\bar{Y}C_1{}^2\theta}{(1+N)}+2k_2k_1\bar{X}\,\bar{Y}\theta\left\{\frac{C_1{}^2}{(1+N)}-C_1C_0\right\} \qquad (2.14)$$

where, $k_1=\dfrac{A_4A_5-2\bar{Y}^2(1+A_2)A_3}{2A_5^2-2A_3\bar{Y}^2(1+A_1)}$, $k_2=\dfrac{\bar{Y}^2\{2A_5(1+A_2)-A_4(1+A_1)\}}{2A_5^2-2A_3\bar{Y}^2(1+A_1)}$

and $A_1=\theta\left\{C_0{}^2+\dfrac{C_1{}^2}{(1+N)^2}-\dfrac{2C_0C_1}{(1+N)}\right\}$, $\qquad A_2=\theta\left\{\dfrac{3C_1{}^2}{8(1+N)^2}-\dfrac{C_0C_1}{2(1+N)}\right\}$,

$A_3=\bar{X}^2\theta C_1{}^2$, $\qquad A_4=\bar{Y}\bar{X}\theta\dfrac{C_1{}^2}{(1+N)}$, $\qquad A_5=\bar{Y}\bar{X}\theta\left\{\dfrac{C_1{}^2}{(1+N)}-C_0C_1\right\}$.

## 3. Empirical Study

In the support of theoretical results, we have considered the data given in Murthy (1967, pp. 131-132). These data are related to the length and timber volume for ten blocks of the black's mountain experimental forest.

The value of intraclass correlation coefficients $\rho_X$ and $\rho_Y$ have been given approximately equal by Murthy (1967, p. 149) and Kushwaha and Singh (1989) for the systematic sample of size 16 by enumerating all possible systematic samples after arranging the data in ascending order of strip length.

The particulars of the population are given below:

$N = 176$, $\quad n = 16$, $\quad \bar{Y} = 282.6136$, $\quad \bar{X} = 6.9943$,

$S_Y^2 = 24114.6700$, $\quad S_X^2 = 8.7600$, $\quad \rho = 0.8710$.

**Table 3.1** MSE and variances of estimators

| Estimators | PRE $(\bar{y}^*)$ |
|---|---|
| $\bar{y}^*$ | 100.00 |
| $t_1$ | 397.5 |
| $t_2$ | 34.07 |
| $t_3$ | 209.43 |
| $t_4$ | 468.68 |

## 4. Conclusion

In this paper, we have adapted Bahl and Tuteja (1991) estimator in Systematic sampling. It is observed from the Table 3.1 that the proposed estimator $t_4$ under optimum condition performs better than the Swain (1964) estimator $t_1$ and Shukla (1971) estimator $t_2$ and $t_3$.